# Local electrical control of magnetic order and orientation by ferroelastic domain arrangements just above room temperature

L. C. Phillips[1,†,*], R. O. Cherifi[1,†], V. Ivanovskaya[1,†,**], A. Zobelli[2], I. C. Infante[3], E. Jacquet[1], N. Guiblin[3], A. A. Ünal[4], F. Kronast[4], B. Dkhil[3], A. Barthélémy[1], M. Bibes[1] and S. Valencia[4,***]

[1]Unité Mixte de Physique CNRS/Thales, 1 av. Fresnel, 91767 Palaiseau & Université Paris-Sud, Orsay 91405, France.

[2]Laboratoire de Physique des Solides, Université Paris-Sud, CNRS UMR 8502, Orsay 91405, France.

[3]Laboratoire SPMS, UMR 8580, Ecole Centrale Paris-CNRS, Grande voie des vignes, Châtenay-Malabry 92290, France.

[4]Helmholtz-Zentrum Berlin für Materialen und Energie, Albert-Einstein-Strasse 15, Berlin 12489, Germany.

[†]These authors contributed equally to this work.
[*]e-mail: lee.phillips@thalesgroup.com
[**]e-mail: v.ivanovskaya@gmail.com
[***]e-mail: sergio.valencia@helmholtz-berlin.de

**Abstract**

Ferroic materials (ferromagnetic, ferroelectric, ferroelastic) usually divide into domains with different orientations of their order parameter. Coupling different ferroic systems creates new functionalities, for instance the electrical control of macroscopic magnetic properties including magnetization and coercive field. Here we show that ferroelastic domains can be used to control both magnetic order and magnetization direction at the nanoscale with a voltage. We use element-specific x-ray imaging to map the magnetic domains as a function of temperature and voltage in epitaxial FeRh on ferroelastic $BaTiO_3$. Exploiting the nanoscale phase-separation of FeRh, we locally interconvert between ferromagnetism and antiferromagnetism with a small electric field just above room temperature. Our results emphasize the importance of nanoscale ferroic domain structure to achieve enhanced coupling in artificial multiferroics.

**Introduction**

Electrical control of magnetic properties[1,2] is an important research goal for low-power write operations in spintronic data storage and logic[3]. Previously, interfacial strain/charge effects have been used to effect electrical control of magnetic anisotropy[4,5], domain structure[6,7], spin polarization[8,9] and exchange bias[10]. Magnetic order has also been electrically controlled[11], but demonstrations have been restricted to interchanging ferromagnetism and paramagnetism via small changes in the Curie temperature $T_C$. Such changes only work in a small temperature range near $T_C$.

We recently controlled magnetic order between antiferromagnetism and ferromagnetism above room temperature, by driving the first-order magneto-structural transition of a $Fe_{50}Rh_{50}$ (FeRh) thin film grown on ferroelectric/ferroelastic $BaTiO_3$ (BTO), creating a very large magnetoelectric (ME) effect[12]. The FeRh transition is interesting because first-order transitions are rare in metals, because one can drive the transition by temperature[13], pressure[14] or magnetic field[15], and because it exhibits phase separation over a range of conditions. Recent imaging studies[16-18] showed that the FeRh transition proceeds by the formation of robustly ferromagnetic sub-micron-sized regions that subsequently grow and coalesce.

Here we use magnetic imaging by photoelectron emission microscopy with X-ray excitation (XPEEM) and X-ray magnetic circular dichroism (XMCD) contrast to reveal that the ME effect in FeRh on BTO is driven by the creation and annihilation of the same regions, due to coupling to ~10 µm-wide BTO ferroelastic domains. The two domain structures are easily distinguished due to their different length scales. Ab-initio calculations confirm the enhanced relative stability of antiferromagnetic FeRh on BTO $c$-domains compared to $a$-domains, suggesting that local ME effects are driven by strain. The local ME effects are greater than our previously observed macroscopic ME effects, and amount to complete local interconversion between antiferromagnetism and ferromagnetism.

**Experiment concept**

We deposit and anneal our FeRh film at high temperature (see Methods), where BTO is in the non-ferroelectric cubic (**C**) phase with lattice constant $a_\mathrm{C}$. FeRh grows with the ordered body-centred cubic (bcc) CsCl-type structure and ferromagnetic (FM) order. The FeRh grows unstrained, but highly aligned with $[100]_\mathrm{FeRh}$ ∥ $[110]_\mathrm{BTO}$ and $[001]_\mathrm{FeRh}$ ∥ $[001]_\mathrm{BTO}$. On cooling, both components undergo non-reconstructive first-order phase transitions. BTO transforms sharply at the ferroelectric Curie temperature $T_\mathrm{C}$ ~ 400 K to the ferroelectric tetragonal (**T**) phase with parameters $a_\mathrm{T} < c_\mathrm{T}$ and ferroelectric polarization $P$ ∥ $[001]_\mathrm{T}$ [Fig. 1(a)]. If no external electric field is applied, the **T**-BTO divides into ferroelastic domains separated by 90° walls, each subdivided into non-ferroelastic ferroelectric domains separated by 180° walls[19]. On the $(001)_\mathrm{C}$-oriented top surface of BTO, one sees ferroelastic $c$-domains ($a$-domains) with polarization perpendicular (parallel) to the surface and a square (rectangular) unit cell projection with surface area $a_\mathrm{T}^2$ ($a_\mathrm{T}c_\mathrm{T}$) per unit cell. The balance of $a$-domains and $c$-domains can be altered by applying a voltage $V$ across the thickness of BTO. Several domain arrangements are observed experimentally[5,6] depending on the thermal and electrical history of the BTO [Fig. 1(b)]. For example, cooling through $T_\mathrm{C}$ with no applied voltage across BTO yields a **T**-BTO texture with mostly *a*-domains separated by 90° domain walls. On the other hand, cooling through $T_C$ in the presence of a voltage, or applying voltage isothermally below $T_C$, yields an almost fully *c*-domain state. Removing an applied voltage isothermally below $T_C$ creates small *a*-domains separated from *c*-domains by buried 90° domain walls. The creation of *a*-domains may be driven by the elastic energy of in-plane-aligned defect dipoles[20].

FeRh transforms gradually between ~420 K and ~360 K via a wide range of phase separation, to a phase with antiferromagnetic (AF) order and the same cubic lattice, albeit with a ~0.3% smaller lattice constant[14]. The mechanical coupling of FeRh and BTO creates interplay between the two transitions. On **C**-BTO [Fig. 1(c)], FM-FeRh experiences only a small biaxial strain from thermal expansion mismatch [-0.2% on cooling from the deposition temperature of 903 K to the BTO $T_C$ of 400 K][21,22]. On **T**-BTO, FM-FeRh experiences either a further biaxial compression on *c*-domains [Fig. 1(d)] or a monoclinic distortion on *a*-domains [Fig. 1(e)]. As the greater compressive strain of *c*-domains shifts the AF-FM transition to higher temperatures [Fig. 1(f)], large ME effects [Fig. 1(g)] are expected if we first heat (cool) from low (high) temperature on BTO consisting of *c*-domains (*a*-domains), and transform isothermally towards *a*-domains (*c*-domains) by changing the voltage across BTO.

**Results**

We studied FeRh on BTO in XPEEM (see Methods), controlling the sample temperature *T* and the voltage *V* across BTO. Tuning the incident X-ray energy, we find the expected absorption peaks for Fe and Rh [Fig. 2(a)]. The peak position and shape imply that both are in the metallic state. A small $FeO_x$ shoulder at 710 eV develops only after irradiation for several days, so our film is essentially stable in our measurement conditions. XMCD images at Fe and Rh edges [Fig. 2(b)] reveal identical magnetic maps over a 20-µm field of view, confirming the presence of the FM phase with the Fe & Rh moments being parallel. Hereafter we will show images at the Fe $L_3$ edge (706.8 eV) whose dichroism is larger[23]. Blue (red) regions with positive (negative) XMCD are FM phase with an in-plane magnetization component parallel (antiparallel) to the incoming X-rays. White regions with zero XMCD are AF phase, or FM phase with magnetization perpendicular to the incoming X-rays. We see phase separation on a ~1µm length scale, which is longer than previous reports[16], probably due to different thickness and strain in our sample. The third phase at FM-AF interfaces[18], in which Rh magnetism is supposed to exceed Fe magnetism, is not observed. The relatively large phase separation scale implies a small interfacial phase volume fraction.

We access the four BTO domain arrangements of Fig. 1(b) by appropriately changing $T$ and $V$. XMCD images at azimuthal angles of 0° and 90° reveal how FeRh responds to changes in the substrate [Fig. 3]. On spatially homogeneous BTO states i.e. **C**-BTO or a single large $c$-domain of **T**-BTO, we see a FeRh microstructure consisting of 1 µm-wide equiaxed FM & AF regions. On inhomogeneous BTO states e.g. mixed $a_1$-$a_2$ or $a$-$c$ domains, we also observe a superstructure of 5-10 µm-wide lamellar regions separated by parallel straight lines. Given that these lamellar structures are very similar to ferroelastic 90° domain walls of BTO (001) crystals[6,21] both in size and in angle with respect to the crystal axes, we identify them as regions of FeRh above different BTO ferroelastic domains. XMCD vector maps [Fig. 3, bottom] reveal that FeRh acts as an indicator film to reveal the underlying BTO domains in two ways: either by alternating magnetization *orientation* on $a_1$-$a_2$ domain patterns, or of ferromagnetic *phase fraction* on $a$-$c$ domain patterns. The changes in FeRh are due either to ferroelastic strain or ferroelectric charge. We exclude charge effects that are permitted within one exchange length[24] of the FeRh/BTO interface, because XPEEM probes the top ~5 nm of our 22-nm film, and therefore cannot probe the charge-affected region. We conclude that FeRh is affected by dynamic strain from ferroelastic BTO domains which extends through the sample thickness.

In order to observe local dynamic electric-field-induced changes in FeRh, we performed an isothermal voltage study as sketched in Fig. 1(g), similar to our previous macroscopic measurement[12]. We heat from ~300 K to the measurement $T$ of 385 K with a large applied voltage $V = +150$ V, preserving a $c$-domain BTO state [Fig. 1(b)] and a mostly AF FeRh state [Fig. 1(g)] with negligible XMCD. Reducing the voltage to zero, we create

BTO *a*-domains which propagate slowly across the field of view, transforming FeRh to a mixed state with ~40-60% FM phase [Fig. 4]. Applying a large voltage in the opposite direction, we reconvert the *a*-domains to *c*-domains and partly reverse the FeRh transformation. Some FM-FeRh remains, just as for our macroscopic measurement[12]. The average XMCD over the images for red and blue domains [Fig. 5(a)] shows a peak at a small negative voltage $V_{pk}$ ~ -15 V. $V_{pk}$ also corresponds to a maximum(minimum) in *a*-(*c*-)domain population [Fig. 5(b)] and a minimum in the FeRh out-of-plane lattice constant [Fig. 5(c)] that were measured simultaneously in high-*T* X-ray diffraction during a similar voltage study of a nominally identical sample at 385 K. These measurements provide further proof that the magnetic changes in FeRh are driven by strain arising from interconversion of BTO *a*- and *c*-domains.

**Discussion**

We have observed and dynamically altered two properties of FeRh at a local scale: the magnetization orientation in the FM phase, and the AF/FM phase balance. We have thus demonstrated simultaneous electrical control of magnetic order and magnetic anisotropy at the micron and sub-micron length scales. The changes of magnetization orientation arise from ME coupling, specifically the *anisotropic* part of magnetostriction $\lambda_t$, which was previously discounted as negligible[24,25]. Our XPEEM images on $a_1$-$a_2$ BTO domains show that $\lambda_t$ is non-zero and positive. The changes of phase fraction arise from relative changes in the free energies of AF and FM phases. The phase fraction changes we have seen in XPEEM are larger and more reversible than our previous macroscopic measurements[12] suggest. This may be because: local changes in our XPEEM field of view can exceed global changes over the whole film; we use a larger voltage to bring BTO to electrical saturation; the out-of-plane FM component is hardly detected in XPEEM; and because our macroscopic measurements were performed in a large saturating magnetic field $H$ = 2 T, which can in principle change the dynamics of the transition.

To understand the link between imposed strain and the relative stability of AF and FM phases, we perform first-principles calculations in the framework of density-functional theory (see Methods). In a previous work, we compared the effect of hydrostatic "isotropic" and biaxial tetragonal "anisotropic" strain in FeRh[12], and showed that the AF phase is favoured by compressive distortions of both types. Here we consider the additional effect of transforming FeRh from tetragonal to monoclinic via in-plane shear. We impose values of the in-plane parameter $a_{FeRh}$ and monoclinic angle $\gamma$ [Fig. 1(e)] and allow the out-of-plane parameter to relax. In this way, we explore possible strain states of FeRh on $a$- and $c$-domains of **T**-BTO.

Calculations of internal energy for AF-FeRh [Fig. 6(a)] and FM-FeRh [Fig. 6(b)] reveal that for all angles $\gamma$, compressive strain causes a steep rise in energy in FM-FeRh, and a weaker rise in energy in AF-FeRh, as we saw previously[12] for $\gamma = 90°$. In contrast, the two phases respond similarly to changes in $\gamma$. The zero-$T$ internal energy difference $\Delta E$ between FM and AF phases for the same strain and $\gamma$ [Fig. 6(c)] is correlated with the AF-FM transition temperature, with a larger $\Delta E$ implying a higher transition temperature[12]. We find that $\Delta E$ is sensitive to the strain via the imposed $a_{FeRh}$, but is almost independent of $\gamma$. For the estimated strain states of FeRh on BTO in our XPEEM study [markers, Fig 6(c)], $\Delta E$ is larger on BTO $c$-domains than on $a$-domains at the same $T$. As the AF phase is favoured on BTO $c$-domains, we anticipate higher (lower) transition temperatures on BTO $c$-domains ($a$-domains), in agreement with our experiments.

**Conclusions**

In summary, we have directly imaged complete local interconversion between antiferromagnetic and ferromagnetic phases of FeRh just above room temperature in response to voltage across the BTO substrate. The effects persist through the thickness of the film. Ab-initio calculations confirm that the ME effects are driven by local strain due to the enhanced AF-FeRh stability on ferroelastic BTO $c$-domains.

Our large local ME coupling could be further increased in thicker FeRh films where the metamagnetic transition can be sharper[26], or by working with Pd-doped FeRh[27] close to room temperature where the BTO tetragonality $c/a$ and associated strains are larger. More broadly, our work shows that giant changes in functional properties can be achieved with moderate stimuli by combining ferroic domain structures with phase-separated materials with first-order transitions.

## Methods

### Sample preparation

22-nm-thick FeRh thin films were deposited on a $5 \times 5 \times 0.5$-mm$^3$ BTO (001) crystal (SurfaceNet GmbH) by radiofrequency sputtering at 903 K with a power of 45 W and in argon pressure of 0.860 Pa. The films were then annealed *in situ* at 1003 K for 90 min. To apply voltage across BTO we use FeRh for the top electrode, and a layer of conducting silver paste or sputtered Pd for the back electrode.

### XPEEM and XMCD measurements.

High-resolution magnetic images were taken at the spin-resolved photo-emission electron microscope at the synchrotron radiation source BESSY II operated by the Helmholtz-Zentrum-Berlin. This set-up is based on an Elmitec III instrument with an energy filter, permanently attached to an undulator beamline with full polarization control, in an energy range from 80 to 2,000 eV. The lateral resolution of the spin-resolved photo-emission electron microscope is about 25 nm for X-ray excitation. For magnetic imaging the photon energy was tuned to the $L_3$ resonance of iron (706.8 eV), exploiting the element-specific XMCD. Each of the XMCD images shown was calculated from a sequence of images taken with circular polarization (90% of circular photon polarization) and alternating helicity. After normalization to a bright-field image, the sequence was drift-corrected, and frames recorded at the same photon energy and polarization have been averaged. The Fe magnetic contrast is shown as the difference of the two average images with opposite helicity, divided by their sum. The magnetic contrast represents the magnetization component pointing along the incidence direction of the X-ray beam, and we study different in-plane components by rotating the sample azimuth in the veam. Vector maps were created by combining two XMCD images with the incident beam along two perpendicular in-plane directions, after correction for drift and distortion.

## High-temperature X-ray diffraction

We used a home-made two axis X-ray diffractometer in the Bragg-Brentano geometry with Cu-$K\alpha_1$ radiation emitted by a 18 kW rotating anode (Rigaku). A designed sample holder allows in-situ x-ray measurements by varying simultaneously the temperature and the applied voltage.

## First-principles calculations

We use a plane-wave pseudopotential approach as implemented in the Quantum ESPRESSO package[28]. The exchange- correlation functional is treated in the Perdew- Burke- Ernzerhof revised for solids (PBEsol) generalized gradient approximation. Atomic relaxation calculations were performed using a shifted $8 \times 8 \times 8$ Monkhorst-Pack grid for $k$-point sampling and an energy cutoff of 80 Ry for the plane-wave basis. Atomic positions were converged until the Hellmann-Feynman forces on each atom became less than 20 meV/Å. Simulated AF and FM phases of bcc FeRh have equilibrium cell parameters of 2.95 Å and 2.97 Å in a good agreement with experiments.[14] Similarly to previous ab-initio studies[29,12], the AF phase is more stable by 45meV/atom. The estimated bulk moduli for AF and FM phases are 227 and 222 GPa, respectively, agreeing well with previous theoretical calculations ([29] and references therein) and consistent with experimental estimates[30].

## Acknowledgements

We are very grateful to S. Hämäläinen, S. van Dijken and A. Mougin for magneto-optical characterization. This work received financial support from the French Agence Nationale de la Recherche through project NOMILOPS (ANR-11-BS10-0016) and the European Research Council Advanced Grant FEMMES (contract no. 267579). R.O.C. acknowledges financial support by Thales through a CIFRE PhD grant.

**Author contributions**

M.B. and A.B. initiated the study. A.B., R.O.C. and S.V. conceived the experiments. R.O.C. prepared the samples with the assistance of E.J. L.C.P., A.A.Ü., S.V. and F.K. characterized the sample by XPEEM. L.C.P., S.V. and F.K. analyzed the XPEEM data. L.C.P., I.C.I., B.D., N.G. and R.O.C. carried out the X-ray diffraction experiments. V.I. and A.Z. performed the first-principles calculations. L.C.P. and V.I. wrote the manuscript with input from M.B. and A.B. All authors contributed to the manuscript and the interpretation of the data.

**Additional information**

The authors declare no competing financial interests.

**Figure 1.**

**Experiment concept sketches.**

(a) 3D sketch of the unit cell shape of tetragonal BTO. A white arrow indicates the polarization direction. (b) 2D projection of ferroelastic domains on the BTO $(001)_C$ surface: crystal phase [**C** = cubic; **T** = tetragonal] and tetragonal ferroelastic domain orientations [$(a_1, a_2)$ = $a$-domains; $c$ = $c$-domains]. Arrows indicate operations that convert between the observed domain states: *1* - cool through $T_C$ without voltage *V*; *2* - cool through $T_C$ in applied *V*; *3* - heat through $T_C$; *4* - apply *V*; *5* remove applied *V*. (c-e) Overlaid unit cell shapes of FeRh (green) and BTO (brown) for (c) cubic BTO, and (d) $c$-domains and (e) $a$-domains of tetragonal BTO. (f) Sketch of the expected magnetization (*M*) versus temperature (*T*) for a strained FeRh film on $c$-domains and $a$-domains of tetragonal BTO. (g) Proposed experiments for large magnetoelectric effects: heat (cool) FeRh from low (high) temperature on BTO $c$-domains ($a$-domains), then transform BTO to $a$-domains ($c$-domains) to obtain a large increase (decrease) of magnetization.

**Figure 2.**

**X-ray absorption and XMCD in XPEEM.**

(a) X-ray absorption spectra of FeRh with horizontal linear polarization averaged over a 20-μm field of view, showing Fe $L_{3,2}$ and Rh $M_{3,2}$ peaks. (b) XMCD images of the same area at Fe $L_3$ and Rh $M_3$ edges. Blue (red) areas have a magnetization component parallel (antiparallel) to the incoming X-rays.

**Figure 3.**

**XMCD on various BTO domain arrangements.**

From top to bottom: sketches of expected BTO domains with reference to Fig. 1(b); corresponding Fe-XMCD images at azimuthal angle 0° with incident x-rays along page vertical; Fe-XMCD images at azimuthal angle 90° with incident x-rays along page horizontal; in-plane XMCD vector maps. Black lines mark inferred ferroelastic domain wall locations. Double-headed red-white-blue arrows indicate the component of magnetization that is measured in each row.

**Figure 4.**

**Magnetoelectric effect by XMCD.**

XMCD images at 385 K at several voltages $V$ during a stepwise sweep from +100 V to -100 V. The black arrow indicates the order of images in time. Sketches show the expected BTO domains for large and small $|V|$.

**Figure 5.**

**Voltage dependence of XMCD and structural parameters.**

(a) Points connected by solid lines represent average XMCD asymmetry vs. voltage for red (XMCD < 0) and blue (XMCD > 0) domains in XPEEM images at 385 K. Dashed lines show the same data mirrored about the voltage axis. Some of the source images are also shown in Fig. 4. (b,c) HTXRD data at 385 K in a nominally identical sample, taken during stepwise negative (closed symbols) and positive (open symbols) sweeps of the voltage across BTO: (b) fraction of BTO $c$-domains, from the relative areas of BTO 002 and 200+020 peaks; (c) FeRh out-of-plane lattice parameter. Data were deduced by fitting pseudo-Voigt functions to observed peaks in $\theta/2\theta$ scans.

**Figure 6.**

**First-principles energies of FeRh under strain.**

(a,b) Internal energy variations with strain and $\gamma$ relative to the unstrained ground state, for (a) AF-FeRh and (b) FM-FeRh. Strain is given relative to the unstrained equlibrium FM-FeRh cell parameter $a_{0,\text{FeRh}}^{\text{FM}}$, i.e. strain = $(a_{\text{FeRh}} - a_{0,\text{FeRh}}^{\text{FM}})/a_{0,\text{FeRh}}^{\text{FM}}$. (c) Internal energy differences $\Delta E$ between FM-FeRh and AF-FeRh for the same strain and $\gamma$. Red dots represent the estimated experimental percentagewise strain states of FeRh on BTO $c$-domains and $a$-domains at the three temperatures indicated on the right.

Figure 1

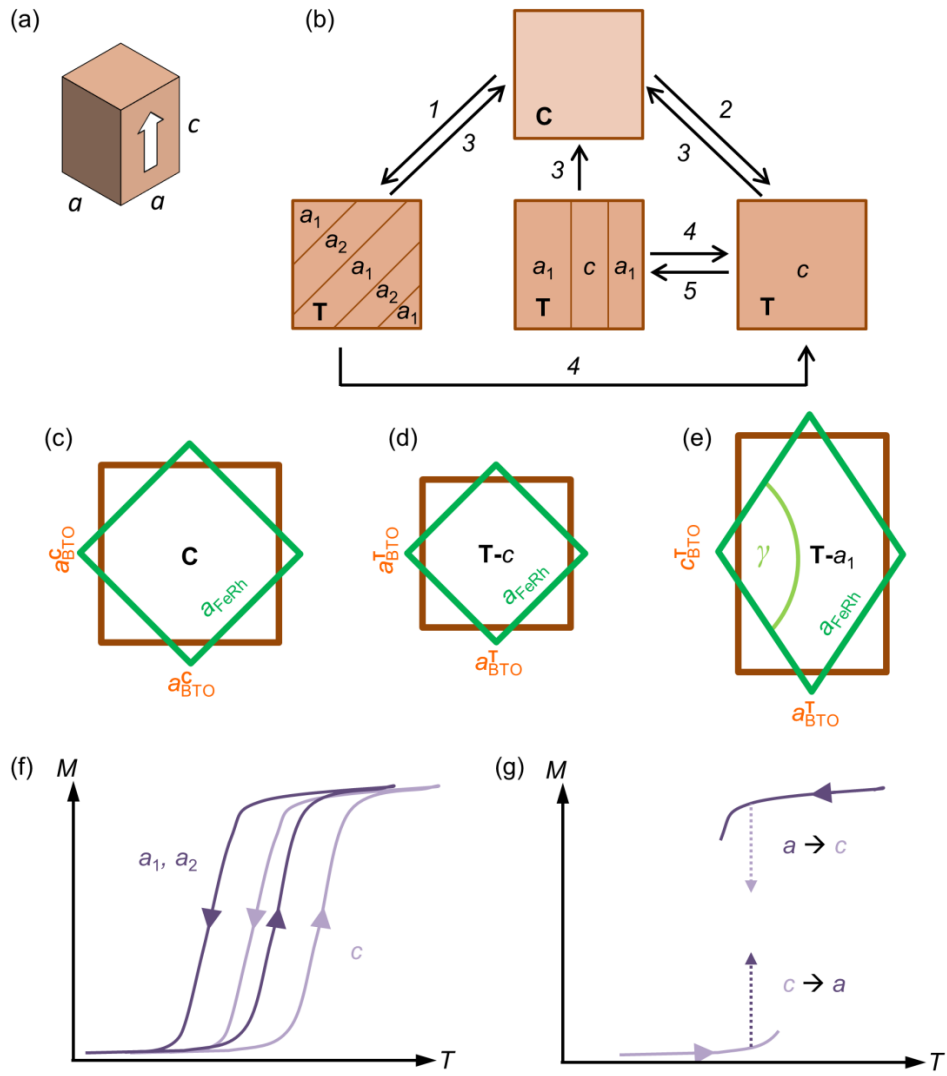

Figure 2

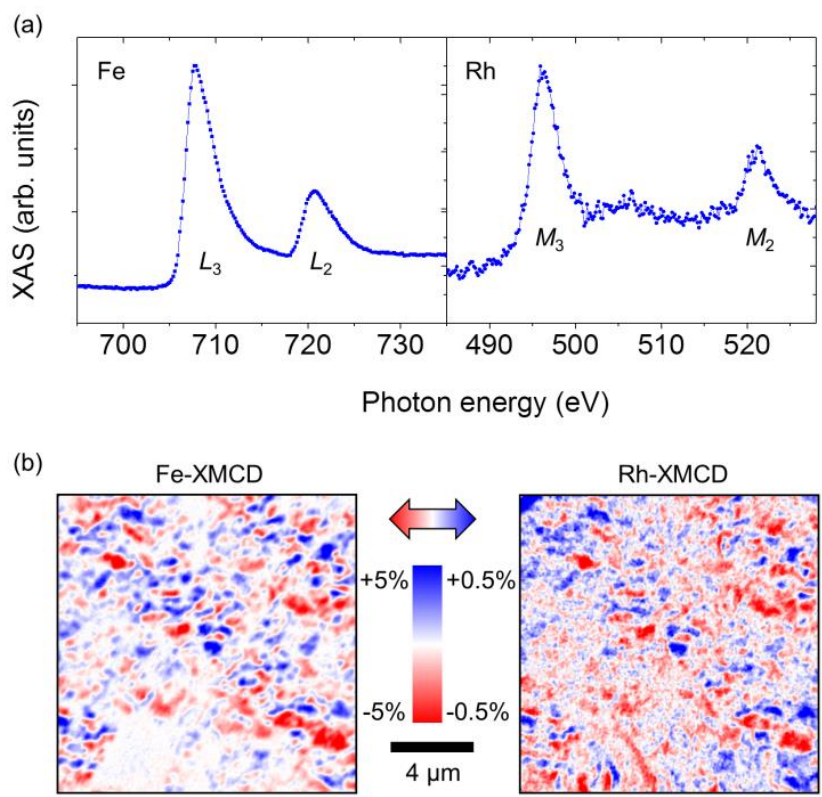

Figure 3

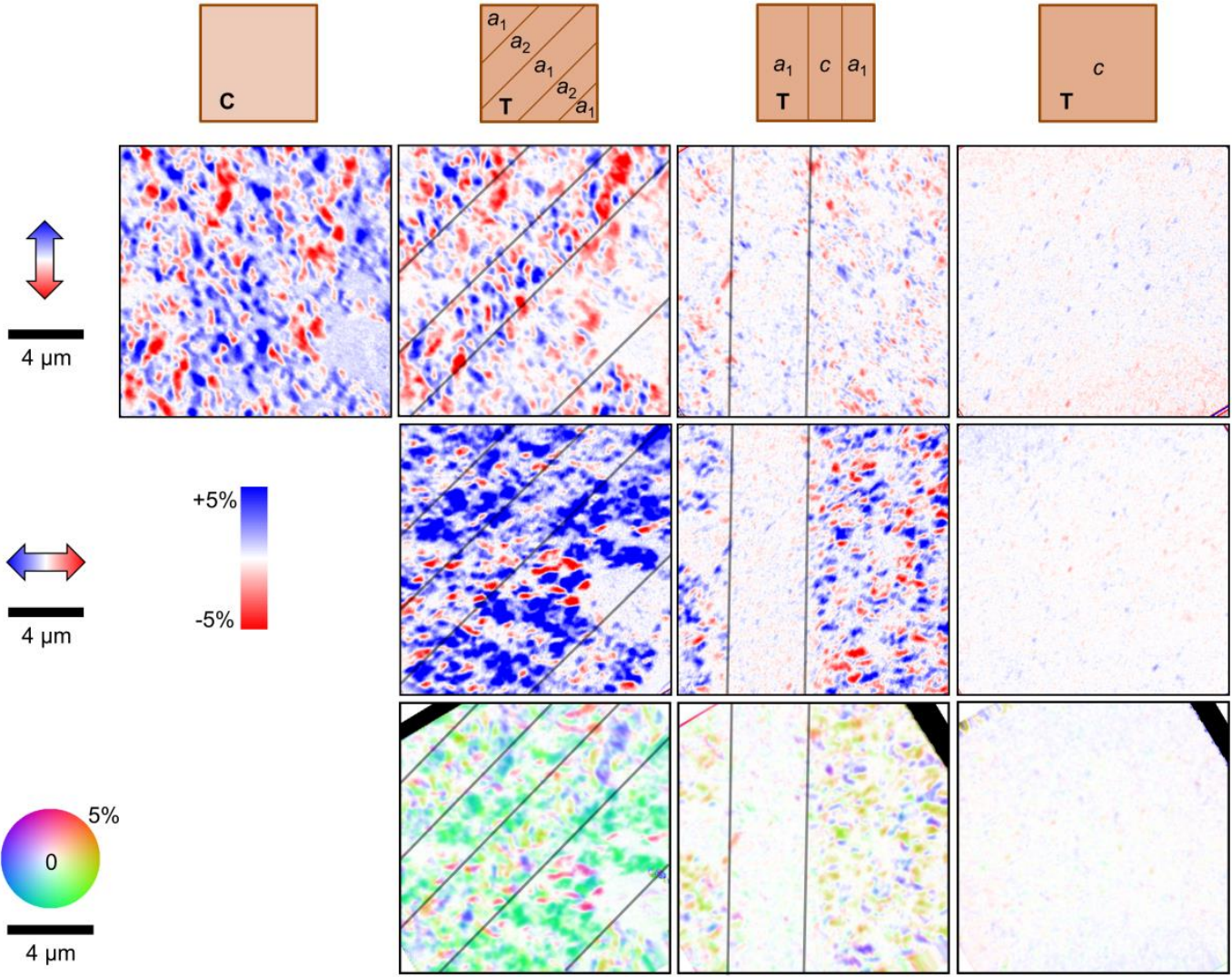

Figure 4

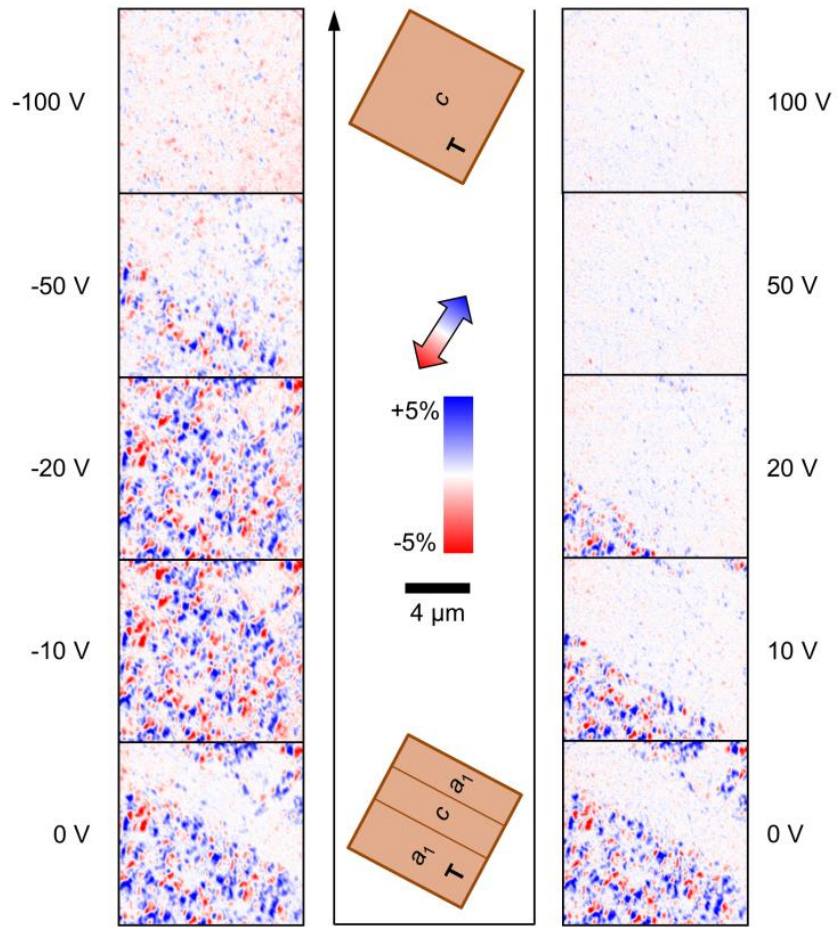

Figure 5

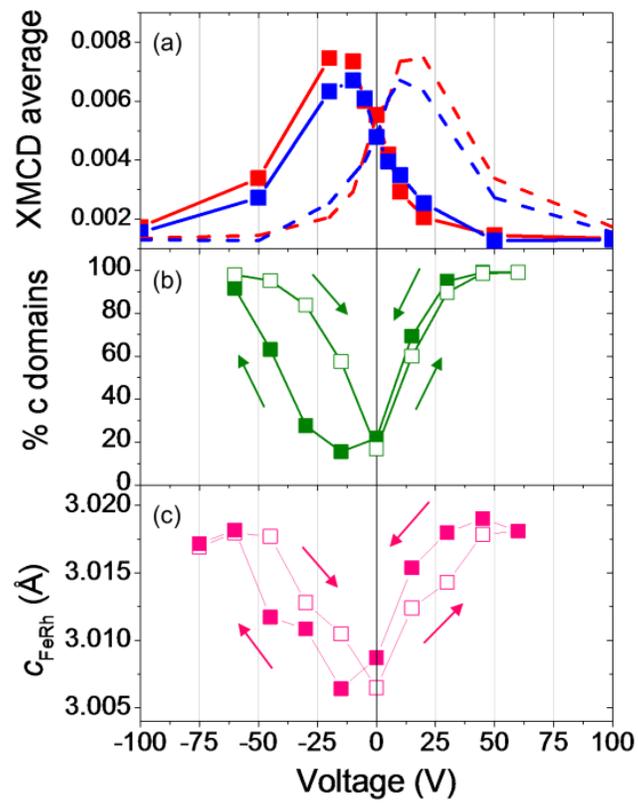

Figure 6

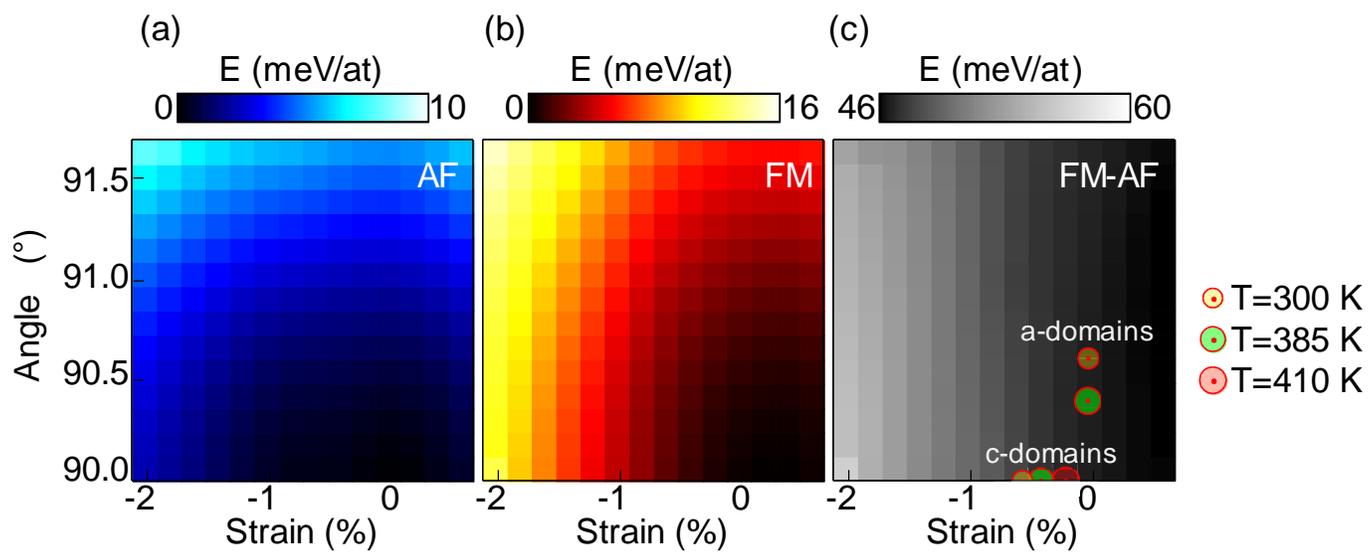